\documentstyle[epsf,rotate]{elsart}

\begin{document}

\begin{frontmatter}
\title{Finite market size as a source of extreme wealth inequality
and market instability}
\author{Zhi-Feng Huang\thanksref{newadd}}
\address{Institute for Theoretical Physics, Cologne University,
D-50923, K\"oln, Germany}
\author{Sorin Solomon}
\address{Racah Institute of Physics, The Hebrew University,
Jerusalem 91904, Israel}
\thanks[newadd]{Present address: Department of Physics, University 
of Toronto, 60 St. George Street, Toronto, ON M5S 1A7, Canada; 
{\em E-mail address}: zfh@physics.utoronto.ca}
\date{}
\maketitle
\begin{abstract}
We study the finite-size effects in some scaling systems,
and show that the finite number of agents $N$ leads
to a cut-off in the upper value of the Pareto law for the
relative individual wealth. The exponent $\alpha$ of the 
Pareto law obtained in stochastic multiplicative market models 
is crucially affected by the fact that $N$ is always finite in real 
systems. We show that any finite value of $N$ leads to properties
which can differ crucially from the naive theoretical results 
obtained by assuming an infinite $N$. In particular, finite $N$ 
may cause in the absence of an appropriate social policy
extreme wealth inequality $\alpha < 1$ and market instability.
\end{abstract}
\end{frontmatter}

Power laws appear ubiquitously as probability distributions
in a wide range of systems. They have been studied for a long time
both theoretically and practically. For instance, the relative
number of people having a wealth between $w$ and $w+dw$
has been found already $100$ years ago to fulfill the Pareto
power law \cite{Pareto}:
\begin{equation}
P(w) dw \sim w^{-1-\alpha} dw.
\label{eq-power}
\end{equation}
From the very beginning it turned out that especially for
small values of exponents ($\alpha < 2$) which imply slow decay
of the distribution at infinity, this distribution form presents
quite a few surprises, paradoxes and unexpected properties.
In particular, the limit of an infinite number of individuals
$N \rightarrow \infty$ is not defined for many of the measurements.

In the present paper we study some dynamical processes where the
naive theoretical results obtained for $N = \infty$ fail for any
arbitrarily large finite $N$. One of the main sources of this 
non-uniform limit is the upper cut-off effect which any finite 
value of $N$ implies. We mean by "the upper cut-off effect" the 
following simple fact: Suppose the wealth of individuals in a 
particular system is distributed by a power law with $\alpha < 2$.
Then, for any value $w$, there is a finite probability that an 
individual $i$ possesses a wealth $w_i$ larger than $w$. For a 
finite system, if one defines by $x_i = w_i / W$ the relative 
wealth of the individual $i$, with $W$ the total wealth, one has 
obviously probability $0$ that any $x_i$ will take a value larger 
than $1$. Therefore, obviously, the distribution of the relative 
wealth $x_i$ cannot be a power law and it has an upper cut-off. 
Moreover, when this cut-off is taken into account the very value 
of the exponent $\alpha$ (in the region of $x_i$ values in which 
the power law does apply) is affected.

This discontinuous behavior of the limit $N \rightarrow \infty$
is important in practical applications where the samples are always 
finite. Indeed, it means that certain theoretical results obtained 
within the assumption of $N =\infty$ are invalid. The problem of
finite-size effects in stock market models has been investigated
\cite{Stauffer99}, especially for the price changes. Here we 
concentrate mainly on the effect of finite-size cut-off on the 
power-law of the wealth distribution.

In this paper we highlight the qualitative difference between the 
behaviors of the finite $N$ system and the theoretical results for 
$N= \infty$. We consider first the example of a pure multiplicative 
stochastic process that leads for $N= \infty$ to the log-normal 
distribution. The log-normal distribution implies an effective 
exponent which at asymptotically large values diverges to 
$\alpha \rightarrow \infty$. This is compared with the actual 
behavior for any finite $N$ which converges towards $\alpha =0$.
Then, we study the multiplicative stochastic processes with a lower
bound, which lead theoretically for $N= \infty$ to power laws
with $\alpha > 1$ \cite{Ijiri-Simon,Solomon96}. We contrast it 
with the finite $N$ systems where values $\alpha < 1$ can take 
place.

In particular in financial systems, our results imply that in the 
absence of a social security policy which takes appropriately
into account the effects of a finite population $N$, the Pareto
exponent $\alpha$ may fall to values $\alpha < 1$. This in turn
means the concentration of an $O(1)$ fraction of the total wealth
in the hands of just a few individuals even if the market size is
asymptotically large.
We show that beyond the moral and social problems which such an
extreme unequal distribution of wealth implies, $\alpha < 1$ also
means that the financial markets themselves become unstable as their
fluctuations become macroscopic.

Let us start with a multiplicative random walk process
involving $N$ independent variables (or individual wealths
in financial system) $w_i$, $i=1,\dots,N$, which
undergo each the updating process:
\begin{equation}
w_i(t+1) = \lambda_i (t) w_i(t),
\label{eq-wi}
\end{equation}
where the random numbers $\lambda_i$ are independent but extracted
with the same time-fixed and $i$-independent probability distribution
$\Pi(\lambda)$.

According to the generalized central limit theorem, if the
distribution $\Pi(\lambda)$ is characterized by the mean
\begin{equation}
\langle \ln \lambda \rangle = v,
\label{eq-v}
\end{equation}
and variance
\begin{equation}
\langle (\ln \lambda)^2 \rangle - \langle \ln \lambda \rangle^2 = D,
\label{eq-D}
\end{equation}
the asymptotic time dependence of the $w$ probability distribution
is:
\begin{equation}
P(w,t) dw= {1 \over \sqrt{2\pi D t}}
\exp\left(-{1 \over 2Dt} (\ln w -vt)^2 \right) {1 \over w} dw.
\label{eq-lognormal}
\end{equation}
In particular, it was pointed out that this log-normal distribution
can be regarded as an effective power law (\ref{eq-power}) with
exponent $\alpha$ changing with $w$ \cite{Montroll82}. For small
$w$, $\alpha \sim 0$ which corresponds to $1/w$ distribution,
while in the large $w$ tail, this distribution (\ref{eq-lognormal})
behaves as a power law with exponent diverging to infinity
\cite{Sornette97}.

To see this, let us put the distribution in the form \cite{Sornette97}:
\begin{equation}
P(w,t) = {\frac {1}{\sqrt{2\pi D t}}}
\frac {\exp(\alpha(w) vt)}{w^{1+\alpha(w)}},
\label{eq-logpower}
\end{equation}
where the effective $w$-dependent power exponent $\alpha (w)$
is defined by
\begin{equation}
\alpha (w) = {\frac {1}{2 D t}} \ln \left(\frac {w}{\exp(vt)}\right).
\label{eq-alpha}
\end{equation}
Then formally $\alpha (w\rightarrow\infty) = \infty$,
i.e., the tail of the distribution has an effective power law with
exponent infinity.

We will show however that this tail is completely irrelevant for
any finite system at large enough times: there is simply no
variable $w$ with such large value within the system. As the time 
goes on and the distribution spreads over larger intervals, the 
initial number of variables $N$ becomes too sparse to sample the 
far away tail. Thus, in reality, the behavior of the theoretical 
$N= \infty$ distribution at asymptotic times is quite different 
for any system containing a finite number $N$ of variables.

An upper bound for the values which the $w_i$'s can take is given
by the total wealth in the system, and obviously none of the
individuals can have a larger value, that is,
\begin{equation}
w_i < W,
\label{eq-cutoff}
\end{equation}
with the total wealth $W$ defined as
\begin{equation}
W(t) = \sum\limits_{i=1}^{N} w_i(t) = N \bar{w}(t).
\label{eq-wtot}
\end{equation}
Thus, from Eqs.\ (\ref{eq-alpha}) and (\ref{eq-cutoff}) we have
\begin{equation}
\alpha < {\frac {1}{2 D t}} \ln \left(\frac {W(t)}{\exp(vt)}\right).
\label{eq-bound}
\end{equation}

In the limit $N = \infty$, $\bar{w}$ in Eq.\ (\ref{eq-wtot})
is proportional to the average value
\begin{equation}
\langle w \rangle = \int w P(w) dw,
\label{eq-<w>}
\end{equation}
and for the multiplicative system (\ref{eq-wi}) it is \cite{Marsili98}
\begin{equation}
\langle w \rangle = \langle \lambda \rangle ^t
= \exp(\ln \langle \lambda \rangle t).
\label{eq-avew}
\end{equation}
However, the result is very different for finite $N$. There, the
observed mean wealth $\bar{w}=\sum_i w_i /N$ is much less than the
$N= \infty$ average $\langle w \rangle$, since the average values 
(and other statistical properties) of the multiplicative system are 
determined by the extreme events and the most extreme events can 
appear only for large enough $N$ \cite{Redner90}. It has been shown 
in Ref.\ \cite{Redner90} that the mean $\bar{w}$ approaches the value 
of $\langle w \rangle$ (Eq.\ (\ref{eq-avew})) when the system size
$N > \exp(\mu t)$ (with $\mu$ a constant and decided by the details
of the multiplicative process), which at asymptotic times is far
beyond the size of simulations or real (e.g., financial) systems.
In fact, for finite-size multiplicative systems the total
wealth $W(t)=\sum_i w_i(t)$ at large enough times mainly depends on
the largest value among $w_i$ \cite{Marsili98,Solomon00}, that is,
\begin{equation}
W(t) \sim w_{\rm max} = {\rm max}_{i=1}^N w_i(t).
\label{eq-wmax}
\end{equation}
The value of $w_{\rm max}$ can be obtained analytically
for $N \ll \exp(Dt)$, (i.e., for finite $N$ and asymptotic times)
in the case in which the random numbers $\ln \lambda$
are extracted from a Gaussian distribution of zero mean $v$
and finite variance $D$ \cite{Marsili98,Galambos78}:
\begin{equation}
w_{\rm max} \simeq \exp\left( \sqrt{2Dt\ln N} \right)  .
\label{eq-wmaxG}
\end{equation}

Substituting Eqs.\ (\ref{eq-wmax}) and (\ref{eq-wmaxG})
into (\ref{eq-bound}) one obtains that
\begin{equation}
\alpha < \sqrt{\frac {\ln N}{2Dt}}.
\label{eq-alphaB}
\end{equation}
Thus, Eq.\ (\ref{eq-alphaB}) implies that for finite size $N$ and
large enough time $t$, the effective $\alpha$ value is far from being
infinite even for the large $w$ tail. In fact, the exponent $\alpha$
decreases as time takes larger and larger values.

The numerical simulations of the system for various times and $N$
values confirm this prediction. One way to analyze the simulation 
data and to validate the power law is to use a Zipf plot \cite{Zipf}.
In order to make a Zipf plot one re-labels the various values of the
individual wealths $w_i$ existing in the system at a given time
in descending order as follows. One denotes by $w(1)$ the wealth
$w_{\rm max}$ of the richest individual, $w(2)$ the wealth of the 
wealthiest among the remaining individuals, and so on, until one 
reaches the poorest individual whose wealth is denoted by $w(N)$.
With these notations, the power law Eq.\ (\ref{eq-power}) leads to 
the Zipf law:
\begin{equation}
w(n) \sim n^{-1/\alpha}.
\label{eq-zipf}
\end{equation}
Thus, in the log-log plot of $w(n)$ (the Zipf plot) the power law
(\ref{eq-power}) corresponds to a straight line with slope
$-1/\alpha$. Visually, steeper slopes represent smaller exponents
$\alpha$. E.g., the $1/w$ distribution (corresponding to $\alpha=0$)
leads to an infinite slope in the Zipf plot, and conversely, a flat 
and almost horizontal Zipf plot corresponds to 
$\alpha \rightarrow \infty$.

\begin{figure}
\centerline{\epsfxsize=8.5cm{\rotate[r]{\epsfbox{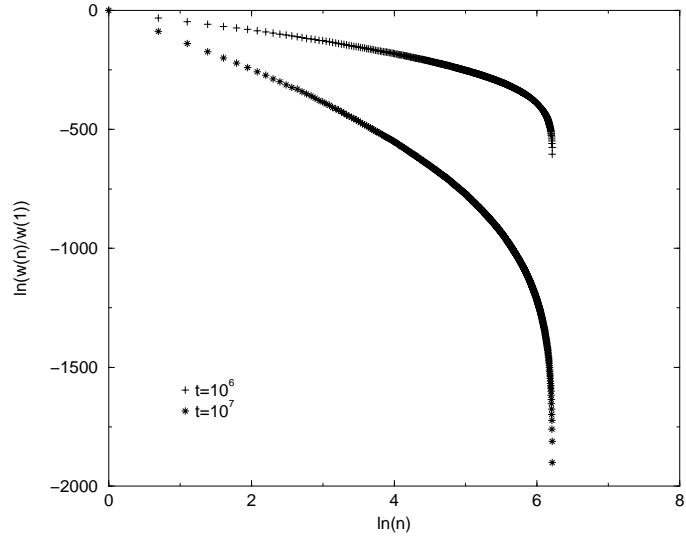}}}}
\caption{The Zipf plots obtained from the simulations of
multiplicative process (\ref{eq-wi}). $\ln \lambda$ is extracted 
from a Gaussian distribution with $v=0$ and $D=0.01$. The system 
size is $N=500$, and the measurements are performed at 2 different
times: $t=10^6$ ($+$) and $10^7$ ($\ast$).}
\label{fig-lnGt}
\end{figure}

\begin{figure}
\centerline{\epsfxsize=8.5cm{\rotate[r]{\epsfbox{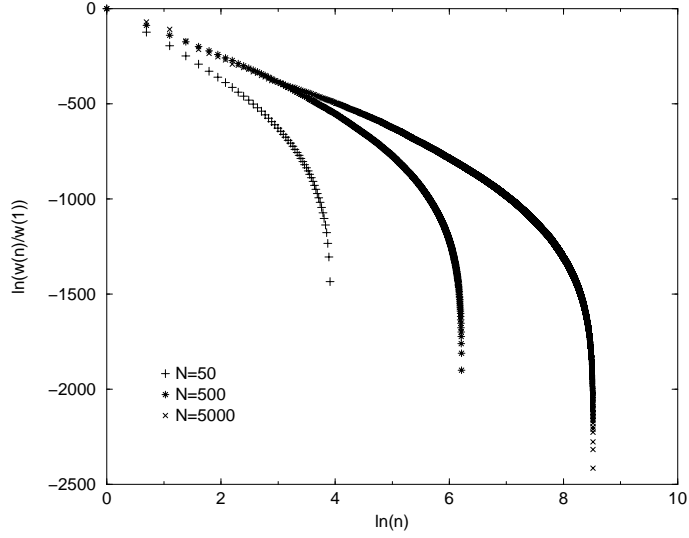}}}}
\caption{The Zipf plots obtained from the simulations of
multiplicative process (\ref{eq-wi}). $\ln \lambda$ is extracted 
from a Gaussian distribution with $v=0$ and $D=0.01$.
The time at which the measurement is performed is
$t=10^7$, and the system sizes are: $N=50$ ($+$), $500$ ($\ast$),
and $5000$ ($\times$).}
\label{fig-lnGN}
\end{figure}

Figs.\ \ref{fig-lnGt} and \ref{fig-lnGN} exhibit the Zipf plots for 
the multiplicative system (\ref{eq-wi}) with $\ln \lambda$ 
distributed by a Gaussian with $v=0$ and $D=0.01$. As expected from 
Eq.\ (\ref{eq-alphaB}), all the slopes for finite $N$ are larger 
than $0$, corresponding to power laws with effective exponents 
$\alpha$ much smaller than the theoretical $N = \infty$ prediction 
$\alpha = \infty$. In fact, in contrast to the $N = \infty$ result,
as time increases $\alpha \rightarrow 0$. This is clearly seen in 
Fig.\ \ref{fig-lnGt} where the slope $-1/\alpha$ becomes steeper with 
increasing time. In accordance with the same Eq.\ (\ref{eq-alphaB}) 
the Zipf plot (Fig.\ \ref{fig-lnGN}) gives steeper and steeper slopes 
($\alpha \rightarrow 0$) as $N$ decreases from $5000$ to $50$.

\begin{figure}
\centerline{\epsfxsize=8.5cm{\rotate[r]{\epsfbox{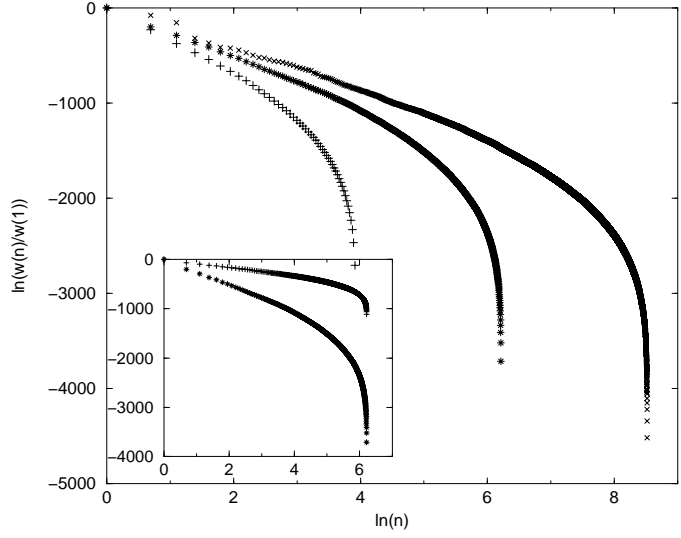}}}}
\caption{The numerical Zipf plots of the multiplicative process
(\ref{eq-wi}). $\ln \lambda$ is extracted from a uniform 
probability distribution between the values $-0.1$ and $0.1$.
The measurements are performed at time $t=1.1\times 10^8$.
The system sizes are $N=50$ ($+$), $500$ ($\ast$), and $5000$ 
($\times$). Inset: $N=500$ and different times: $t=10^7$ ($+$) 
and $1.1\times 10^8$ ($\ast$).}
\label{fig-ln}
\end{figure}

The same results can be obtained for different distributions of
the random factor $\ln \lambda$. Fig.\ \ref{fig-ln} shows the Zipf
plots for $\ln \lambda$ uniformly distributing in a range
$(-0.1,0.1)$. The above phenomena caused by the upper cut-off
(\ref{eq-cutoff}) are still present.

This upper cut-off effect holds for any finite value of $N$ for 
large enough times, and is not limited to the log-normal distribution.
A similar argument insures that other slow-decaying distributions
have sharp truncations and/or departures from the $N=\infty$ results.
In order to uncover these effects one has to use appropriate plots for
each quantity. For instance, the upper cut-off does not affect the 
straight line shape of the Zipf plot for a system obeying the 
power-law distribution (\ref{eq-power}).
However, this cut-off is made explicit by first normalizing
each variable (wealth) to the total wealth and averaging over many
realizations to obtain an estimation of the probability Pareto
distribution $P(w/W)$. By plotting then $\log P(w/W)$ against 
$ \log (w/W)$, obviously one cannot expect a straight line since at 
least for $ w/W > 1$ one has $\log P(w/W) =-\infty$. In the actual 
plots, indeed a dramatic bent is shown in the graph when the variable 
$w$ approaches the total capital $W$.

Thus, we encounter an unexpected subtle effect related with the special
properties of the power laws. The effect originates in the fact
that the density of probability $P(w)$ has to be $0$ at $w= W$ due to 
Eq.\ (\ref{eq-cutoff}), in spite of the naive $N=\infty$ expectation
\cite{Bouchaud-Mezard} that the distribution Eq.\ (\ref{eq-power}) will 
continue to infinity.

A particular model where these observations are relevant
is the multiplicative stochastic model with lower bound
introduced in Refs.\ \cite{Solomon96,Solomon98}. In this model, 
the process (\ref{eq-wi}) is supplemented by a lower bound:
\begin{equation}
w_i(t+1) \geq c \bar{w}(t),
\label{eq-lower}
\end{equation}
(i.e., whenever according to (\ref{eq-wi}) $w_i(t+1)$
becomes less than $c \bar{w}(t)$, the actual updated value
$w_i(t+1)$ is taken as $c \bar{w}(t)$).
Such a lower bound represents in the model the economic effects of 
subsidies, securities, and services.

For infinite $N$, this process leads to the precise power law with
the exponent
\begin{equation}
\alpha = 1/(1-c).
\label{eq-alpha1}
\end{equation}
In reality, that is, for finite $N$, one obtains a different
result \cite{Malcai99}: for $c \ll 1/\ln N$, the exponent is
\begin{equation}
\alpha \cong \ln N / (\ln N - \ln c) < 1,
\label{eq-alpha2}
\end{equation}
while for $c \gg 1/\ln N$ the relation (\ref{eq-alpha1}) is
recovered. (The large $w_i$ scaling properties of this system are
representative of a larger class of systems 
\cite{Solomon96,Bouchaud-Mezard,Solomon98,Malcai99,Solomon99,Bouchaud00,Huang-Solomon00}
of the form:
$$
w_i (t+1) = \lambda_i (t) w_i(t)  + a \bar{w}(t) + 
b(\bar{w},t) w_i (t),
$$
where $b$ is an arbitrary function \cite{Richmond00}.)

\begin{figure}
\centerline{\epsfxsize=8.5cm{\rotate[r]{\epsfbox{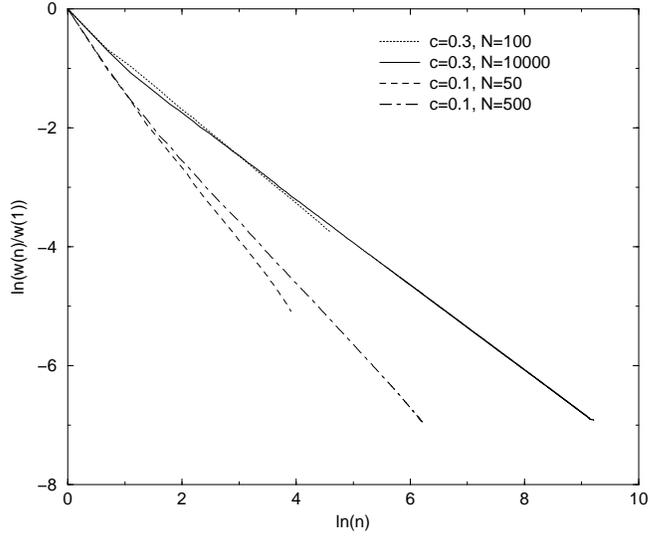}}}}
\caption{The Zipf plots obtained from the numerical simulations
of model (\ref{eq-wi}) with the lower bound (\ref{eq-lower}) and
asynchronous dynamics, for late time $t=2\times 10^5N$ and different
values of $c$ and size $N$. The random factor $\lambda$ is extracted
from an uniform probability distribution between $0.9$ and $1.1$.}
\label{fig-zipf}
\end{figure}

\begin{figure}
\centerline{\epsfxsize=8.5cm{\rotate[r]{\epsfbox{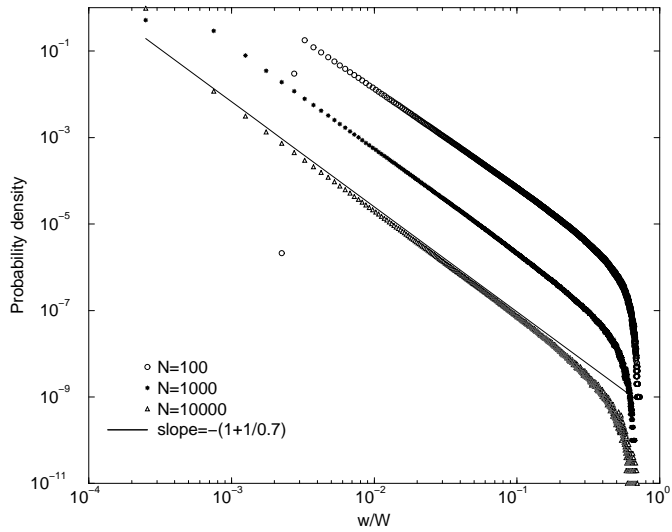}}}}
\caption{Log-log plot of the probability distribution for
normalized wealth $w/W$ of the model with lower bound: Eqs.\
(\ref{eq-wi}) and (\ref{eq-lower}), with simulation parameters:
$c=0.3$, $N=100$, $1000$, and $10000$. $\lambda$ is uniformly
distributed in the range $(0.9,1.1)$. The straight line has the
slope $1+\alpha$, with exponent $\alpha$ determined by
Eq.\ (\ref{eq-alpha1}).}
\label{fig-wi1}
\end{figure}

\begin{figure}
\centerline{\epsfxsize=8.5cm{\rotate[r]{\epsfbox{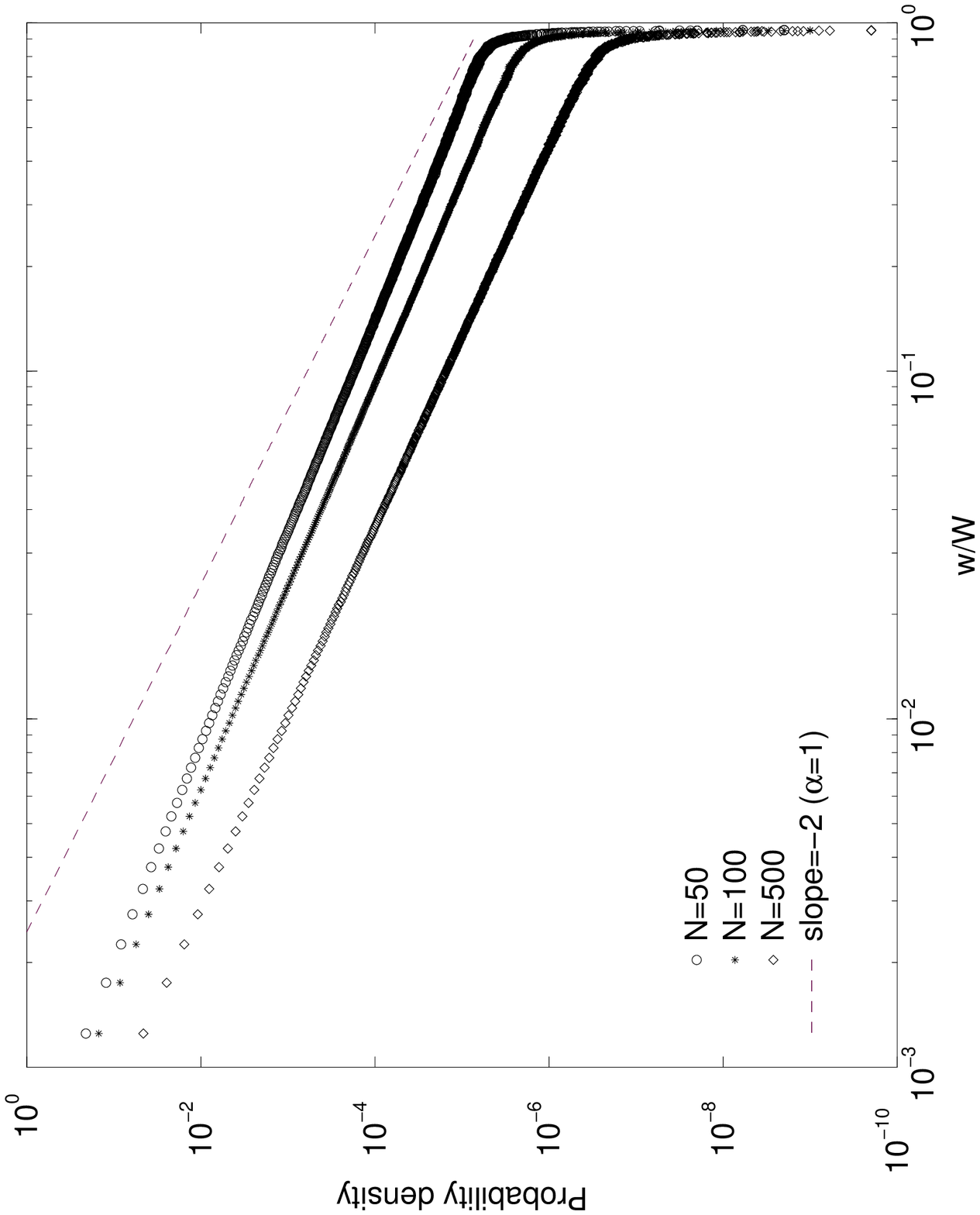}}}}
\caption{Similar to Fig.\ \ref{fig-wi1}, but with parameters
$c=0.05$ and $N=50$, $100$, and $500$, so that $c < 1/\ln N$.
The straight line has the slope $2$, corresponding to
$\alpha=1$.}
\label{fig-wi2}
\end{figure}

Numerical results probing the properties of this
multiplicative stochastic model with lower bound
are shown in Figs.\ \ref{fig-zipf}, \ref{fig-wi1}
and \ref{fig-wi2}. (We used here the realistic asynchronous 
dynamics, i.e., different wealths are updated in successive
random order: at each time $t$ only one wealth is randomly
chosen for updating. However, similar results are obtained for 
the synchronous dynamics). In the simulations the random factor
$\lambda$ is distributed with uniform probability between $0.9$ 
and $1.1$.

As explained above, while the Zipf plots of the individual wealth
configurations do not show any deviation from the precise power-law
behavior (see Fig.\ \ref{fig-zipf}), the Pareto distribution of
the normalized $w/W$, obtained by averaging over many runs, shows a
sharp bent to $\log P(w) = -\infty$ for $w/W \rightarrow 1$
(Figs.\ \ref{fig-wi1} and \ref{fig-wi2}).

In Fig.\ \ref{fig-wi1}, we have $c > 1/\ln N$ and then the formula 
(\ref{eq-alpha1}) holds especially for large $N$. However, when there 
is no significant social security policy, that is, $c < 1/\ln N$ in 
the lower bound Eq.\ (\ref{eq-lower}), one obtains the low values of 
exponent $\alpha<1$, as predicted by Eq.\ (\ref{eq-alpha2}).
This is clearly seen in Fig.\ \ref{fig-wi2}.

For the power-law distribution (\ref{eq-power}), the ratio between 
the largest wealth and the total wealth is 
$w_{\rm max} / W \sim N^{1/\alpha -1}$. Thus, when $\alpha<1$, the 
richest person has an $O(1)$ fraction of the total wealth even if 
the market size is asymptotically large. This extreme financial 
inequality is questionable from moral point of view and unacceptable 
from social point of view \cite{Solomon99,Bouchaud00}.

However, here we make the observation that such an inequality is
equally dangerous even from the narrower point of view of financial
stability. To understand the implications of the $\alpha<1$
on the market fluctuations, recall \cite{Solomon96,Huang-Solomon00}
that the market index fluctuations are expected to also obey 
(modulo the cut-off tails) a power-law behavior with an exponent
close to or larger than that of the individual wealth Pareto
distribution. Telegraphically, the argument in Ref.\ \cite{Solomon96} 
consisted of the following steps:
\begin{itemize}
\item  The market index is identified with the total wealth $W$ in 
the market, i.e. with the sum of the individual invested wealths $w_i$;
\item The market fluctuations over a certain time period are therefore
the sum of the variations of the individual wealths $w_i$, with $i$
randomly selected for updating during that period;
\item The individual variations of the $w_i$'s are, according to Eq.\ 
(\ref{eq-wi}), stochastically proportional to the $w_i$'s themselves,
which are distributed by a power law with exponent $\alpha$;
\item Thus, the market fluctuations are the sums of steps with sizes
distributed by a power law with exponent $\alpha$.
\end{itemize}
Therefore, a wealth distribution with $\alpha<1$ implies market
fluctuations distributed by a power law with a little larger but 
still small effective exponent (modulo the truncation effects,
the details of which are outside the scope of the present paper
and were studied in Ref.\ \cite{Huang-Solomon00}.
They fit well the experimentally observed cross-over to larger
power-law exponents and to exponential-type behavior 
\cite{Huang00,exp00} for extremely large fluctuations.)
In turn this small exponent implies the emergence with finite 
probability of macroscopically large market fluctuations, which 
means that the financial markets in a society with a Pareto
wealth distribution of $\alpha < 1$ are open to a significant
risk of macroscopic market crashes. The conclusion is that a 
responsible social policy which insures the smallest wealth 
$w_{\rm min} \gg {\bar w} / \ln N$ is not just a humane duty
but also a vital interest of the capital markets.

This finite-size cut-off effect also occurs in systems other than
the multiplicative ones, in particular for the critical phenomena.
For the two- or three-dimensional percolation system, which has been
well studied and understood \cite{Stauffer}, the number $n_s$ of
clusters containing $s$ sites each decays for large $s$ as a power
law
\begin{equation}
n_s \sim 1/s^{\tau},
\label{eq-ns}
\end{equation}
in an infinite lattice and at the percolation threshold $p_c$.
The Fisher exponent $\tau$ is $2.05$ for two dimensions ($d=2$),
while for three dimensions ($d=3$) it is $2.19$. In a finite 
lattice, the cut-off behavior similar to Eq.\ (\ref{eq-cutoff}) 
exists, that is, no cluster can be larger than the whole
lattice and then the cluster radius
\begin{equation}
R_s < L,
\label{eq-cutoffd}
\end{equation}
in a $d$-dimensional lattice of linear size $L$, where at
$p_c$ this radius behaves as $R_s \sim s^{1/D}$, with $D=1.9$
for $d=2$ or $2.5$ for $d=3$. The size distributions for
these largest clusters have been investigated \cite{Sen99}.

Thus, things are more complicated for the cut-off effect of the
finite systems. For the critical phenomena like the percolation,
little is known about the extremely rare clusters at $p_c$ with
size $s$ between $L^D$ and $L^d$ which are outside the scaling
region. As in the case of the multiplicative processes such as
Eqs.\ (\ref{eq-wi})-(\ref{eq-lower}), more work is needed in order 
to investigate e.g. the margin of the wealth value obeying the
power-law scaling as well as its dependence on the system size.

Returning to the financial applications, our results caution
against the use of the $N= \infty$ limit in estimating the
parameters of the distributions for wealth and market fluctuations.
In particular, while for $N=\infty$ some random multiplicative
mean field models seem to insure $\alpha > 1$ \cite{Bouchaud-Mezard}, 
in reality for any finite number of individuals $N$, in the absence 
of an appropriate social policy which insures the smallest wealth
$w_{\rm min} \gg \bar{w} / \ln N$ one is led to a very unbalanced
wealth distribution $\alpha <1$. For such a wealth distribution,
a large fraction of the total wealth remains in the hands of
just a few richest individuals. This in turn is bound to lead not 
only to social unrest but also to unbearable financial markets 
fluctuations.

\section*{Acknowledgements}
We thank Dietrich Stauffer for very helpful discussions and
comments. Z.F.H. acknowledges the financial support of SFB 341.

\begin{thebibliography}{}

\bibitem{Pareto} V. Pareto, Cours d'Economique Politique,
Macmillan, Paris, 1897, Vol. 2.

\bibitem{Stauffer99} D. Stauffer, WEHIA e-print,
at http://dibe.unige.it/wehia, June 1999.

\bibitem{Ijiri-Simon} Y. Ijiri and H.A. Simon, Skew 
Distributions and the Sizes of Business Firms, North-Holland, 
Amsterdam, 1977.

\bibitem{Solomon96} M. Levy and S. Solomon, Int. J. Mod. Phys. C
7 (1996) 595; S. Solomon and M. Levy, {\it ibid.} 7 (1996) 745.

\bibitem{Montroll82} E. W. Montroll and M.F. Shlesinger,
Proc. Nat. Acad. Sci. USA 79 (1982) 3380.

\bibitem{Sornette97} D. Sornette and R. Cont, J. Phys. I 7
(1997) 431.

\bibitem{Marsili98} M. Marsili, S. Maslov, and Y.C. Zhang,
Physica A 253 (1998) 403.

\bibitem{Redner90} S. Redner, Am. J. Phys. 58 (1990) 267.

\bibitem{Solomon00} S. Solomon, e-print cond-mat/0008076.

\bibitem{Galambos78} J. Galambos, The Asymptotic Theory
of Extreme Order Statistics, Wiley, New York, 1978.

\bibitem{Zipf} G.K. Zipf, Human Behavior and the Principle of
Least Effort, Addison-Wesley, Cambridge MA, 1949.

\bibitem{Bouchaud-Mezard} J.P. Bouchaud and M. M\'ezard, Physica A
282 (2000) 536.

\bibitem{Solomon98} S. Solomon, in {\it Decision Technologies for
Computational Finance}, edited by A.-P. Refenes, A.N. Burgess,
and J.E. Moody, Kluwer Academic Publishers, 1998.

\bibitem{Malcai99} O. Malcai, O. Biham, and S. Solomon, Phys. Rev.
E 60 (1999) 1299.

\bibitem{Solomon99} S. Solomon, Generalized Lotka Volterra (GLV)
Models, in {\it Applications of Simulation to Social Sciences},
Eds: G. Ballot and G. Weisbuch, Hermes Science Publications, 2000.

\bibitem{Bouchaud00} J.P. Bouchaud, e-print, cond-mat/0008103,
to appear in J. Quan. Finance.

\bibitem{Huang-Solomon00} Z.F. Huang and S. Solomon, e-print,
cond-mat/0008026, Eur. Phys. J. B, in press.

\bibitem{Richmond00} P. Richmond and S. Solomon, e-print,
cond-mat/0010222, J. Quan. Finance, in press.

\bibitem{Huang00} Z.F. Huang, Physica A 287 (2000) 405.

\bibitem{exp00} J. Skjeltorp, Physica A 283 (2000) 486;
J. Masoliver, M. Montero, and J.M. Porr\`{a}, {\it ibid.} 283
(2000) 559.

\bibitem{Stauffer} D. Stauffer and A. Aharony, Introduction
to Percolation Theory, Taylor and Francis, London, 1994.

\bibitem{Sen99} P. Sen, Int. J. Mod. Phys. C 10 (1999) 747;
M.I. Zeifman and D. Ingman, J. Appl. Phys. 88 (2000) 76.

\end{thebibliography}
\end{document}